\documentclass[referee]{aa}
\usepackage{pslatex}
\usepackage{graphics}
\begin{document}
\thesaurus{(	08.02.7;  
		08.04.1;  
		08.15.1;  
		08.16.3;  
		08.22.2;  
		10.07.3)} 
\title{Cluster AgeS Experiment.
CCD photometry of SX~Phoenicis variables in the globular cluster M~55}
\author{Wojtek~Pych \inst{1}, Janusz~Kaluzny \inst{1}, 
Wojtek~Krzeminski \inst{1,2}, A.~Schwarzenberg-Czerny \inst{1,3}
\and Ian~B.~Thompson \inst{2}, 
}
\institute
{Copernicus Astronomical Center, ul. Bartycka 18, 00-716 Warszawa,
Poland
\and
Carnegie Institution of Washington, 813 Santa Barbara Street, Pasadena,
CA 91101
\and
Astronomical Observatory of Adam Mickiewicz University, ul.
Sloneczna 36, 60-286 Poznan, Poland}
\date{Received ................., 2000; Accepted ..............., 2000}
\titlerunning{CASE - SX~Phe variables in M~55}
\authorrunning{Pych et~al.}
\maketitle
\begin{abstract}
We present CCD photometry of SX~Phe variables in the field of the
globular cluster M~55. We have discovered 27 variables, three of which
are probable members of the Sagittarius dwarf galaxy. All of the SX~Phe
stars in M~55 lie in the blue straggler region of the cluster
color-magnitude diagram.  Using  period ratio information we have
identified the radial pulsation modes for one of the observed
variables.  Inspection of the period-luminosity distribution permits
the probable identifications of the pulsation modes for most of the
rest of the stars in the sample.  We have determined the slope of the
period-luminosity relation for SX~Phe stars in M~55 pulsating in the
fundamental mode. Using this relation and the HIPPARCOS data for SX~Phe
itself, we have estimated the apparent distance modulus to M~55 to be
$(m-M)_V=13.86\pm0.25$~mag.
\end{abstract}
\section{Introduction}
The Clusters AgeS Experiment (CASE) is a long term project with a main
goal  of determining accurate ages and distances of globular clusters
by using observations of detached eclipsing binaries (\cite{Paczynski1997}).
As a byproduct we obtain time series photometry
of other variable  stars located in the surveyed clusters. 

M~55 (NGC 6809) is a metal-poor globular cluster in the Galactic halo
($l=9^{\circ}, b=-23^{\circ}$).
Because of its proximity and small reddening ($(m-M)_V=13.76,
E(B-V)=0.07$, \cite{Harris1996}) it was selected as one of the targets in
our survey for eclipsing binaries in globular clusters. 
Olech et~al. (\cite{Olech1999}) presented our investigation of RR~Lyrae 
variables in this cluster. In this contribution we present
the results for the short period pulsating variables.
The relatively large number of SX~Phe variables in M~55
allows us to make a basic statistical analysis of their properties and a
new determination of the slope of the period-luminosity relation.
\section{Observations and data reduction}
In the interval from 1997 May 8/9 to 1997 August 16/17 we carried out
CCD photometry on the 1.0-m Swope telescope at Las Campanas
Observatory.  The telescope was equipped with the SITe3 2k~x~4k CCD
camera with an effective field of view 14.5~x~23~arcmin (2048x3150
pixels were used), at a scale of 0.435~arcsec/pixel.  The cluster was
monitored on 13 nights for a total of 36.4 hours.  The light curves
typically have about 750 data points in Johnson $V$ and about 60 data
points in Johnson $B$.  Exposure times were 150~sec to 300~sec for the
$V$ filter and from 200~sec to 360~sec for the $B$ filter, depending on the
atmospheric transparency and seeing conditions.  On photometric nights
several fields of standard stars (\cite{Landolt1992}) were observed to obtain
transformation coefficients to the photometric standard system.  We
used procedures from the IRAF {\it noao.imred.ccdproc} package for
de-biasing and flat-fielding the raw data. Instrumental photometry was
obtained using DoPHOT (\cite{Schechter1993}).
\section{Light Curves}
\subsection{Identification of variables}
We identified 27 SX~Phe variables in the field of M~55.
Following the nomenclature of Olech et~al. (\cite{Olech1999}) the stars are
designated as NGC~6809~LCO~V16 through NGC~6809~LCO~V42.
Here-after we use the designations V16~--~V42 respectively. Finding
charts for these variables are presented in
Figs. \ref{f1}, \ref{f2} and \ref{f3}.

The positions of these objects in the color-magnitude diagram are shown in
Fig.~\ref{f4}. All of the SX~Phe stars belonging to M~55 lie on the blue
straggler sequence.
Three of the observed SX~Phe type stars: V28, V29, V30,
are 3.5~--~4~mag fainter than the rest of our sample stars.
This difference in magnitude places these 3 stars in the
Sagittarius dwarf galaxy (\cite{Ibata1994}, \cite{Fahlman1996}).
The 24 remaining SX~Phe variables constitute approximately
50~percent  of all the blue straggler stars present in our data.
\subsection{Fourier Analysis }
Preliminary period estimates were obtained using the CLEAN algorithm 
(\cite{Roberts1987}).
We used a method developed by Schwarzenberg-Czerny
(\cite{Schwarzenberg-Czerny1997}) to improve the
period determination and to fit a Fourier series to the $V$-band light curves
in the form:
\begin{eqnarray}
	V = A_o + \sum_{j=1}^k A_j sin(j{\omega}t+\phi_j) \label{e.1}
\end{eqnarray}
where $\omega=2{\pi}/P$ and $P$ is the pulsation period of the star.
The number of harmonics ($k$) was chosen so that the formal errors of 
their amplitudes were smaller than the determined values.
Since the amplitudes of most of the SX~Phe variables in M~55 are smaller 
than 0.1~mag, for 18/27 stars we were able to measure only the base
harmonic ($k=1$). For those stars for which  more harmonics could be
measured we calculated the Fourier parameters: 
\begin{eqnarray}
R_{ij}&=&A_i/A_j \label{e.2} \\
\Phi_{ij}&=&i\Phi_i-j\Phi_i. \label{e.3}
\end{eqnarray}

{\bf In order to help the readers unfamiliar with the AoV periodogram analysis
to appreciate the effects of noise and aliasing on our period analysis
we provide here as an example description of the light variations of 
a double mode pulsating star V41 in the terms of the classical
power spectrum and window function. The window function looks
well as its highest side-lobes due to day and moon cycles do not exceed 83
and 60 percent of the central peak respectively. The window patterns
corresponding to the primary and secondary pulsation frequency in the
respective original and prewhitened power spectrum are little disturbed and
symmetric. Hence our period identifications are unambiguous.
The power spectrum remaining after prewhitening with the two detected
frequencies and their 5 harmonics is rather flat,
consistent with the low frequency calibration errors not exceeding 0.005~mag
and no periodic oscillations with amplitudes exceeding 0.004~mag at
frequencies exceeding 3 c/d. This is consistent with the theoretical
expectations, as any combination amplitudes should be of order of the product
of the detected amplitudes i.e. of 0.001 mag. These values for other stars
remain within factor of 2 of their respective values for V41.

Standard deviation of the residuals is 0.014~mag, consistent with 
that expected for the size of the telescope and stellar magnitude.
Thus observational errors of an average value of 1/4 all observations 
should be as small as $\sigma\sim 0.014/\sqrt{180}\sim 0.001$ mag. However, 
the averages calculated by selecting 1/4 of points around minimum and 
maximum phases should differ by more than the (half)amplitude of the
oscillation, consistent with well over $10\sigma$ significance level
of detection even for the secondary oscillation. The AoV statistics
used by us tends to yield higher significance levels than the above 
simple estimate. }

\subsection{Single mode SX~Phe stars}
The basic parameters derived for the single-mode oscillators are listed
in  Table~\ref{t1}, including the variable number, equatorial
coordinates (J2000.0), derived periods, mean $V$-band magnitudes, mean
$B$-band magnitudes, mean colors ($<$$B$$>-<$$V$$>$), and full amplitudes of
the oscillations in $V$.  Table \ref{t2} presents values of $A_1$,
$R_{21}$, $\Phi_{21}$, $R_{31}$, $\Phi_{31}$  measured for the
single-mode variables.  By analogy to Cepheids, we can look for the
signature of the resonance between the radial pulsation modes in a
$\Phi_{21}$-period  plot (Fig.~\ref{f5}).  This plot suggests
that $\Phi_{21}$ is either constant within the observed period range with a
weighted mean of $2.183 \pm 0.013$ {\bf or slightly increasing
with the period, since according to the Fisher-Snedecor test 
the linear fit $\Phi_{21}=1.954+2.4(\pm1.0)P$
is marginally better than a constant at confidence level 0.95.
Our phases correspond to the sine
Fourier series (Eq.~\ref{e.1}), for the cosine series the
mean phase should be incremented by $\pi/2$. 
This taken into account, both our constant and linear solutions agree 
with \cite{Poretti1999}. Note that our result is based on a much broader
period interval and hence has proportionally stronger weight.} 
The smoothness of the
phase does not reveal the presence of any resonance within the range of
periods observed here. Curiously, within the errors the same phase
difference holds for the principal oscillation of the double mode stars
V31 and V32. These were included in the mean value quoted above.

Fig.~\ref{f6} presents the light curves of the single-mode SX~Phe
variables observed in the field of M~55.
\subsection{Double mode SX~Phe stars}
We constructed  a synthetic light curve for each of the variables using
the measured Fourier parameters for that variable. After subtracting
this light curve from the observed data points, we 
searched for a new period with a new fit of a Fourier series.  If the
full amplitude of the resulting light curve was larger than an
arbitrarily chosen value of 0.01~mag, then the object was classified
as a double mode variable. Two modes of pulsation were detected in the
light curves of 12 of our  variables.  The parameters for these
double-mode SX~Phe variables are listed in Table \ref{t3}, including
the variable number, equatorial coordinates (J2000.0), periods of
pulsations for both modes, mean $V$-band magnitudes, mean $B$-band
magnitudes, mean colors, and full amplitudes in $V$ for the longer
period. Table~\ref{t4} presents the values of $A_1^A$, $R_{21}^A$,
$\Phi_{21}^A$, $A_1^B$ measured for the double-mode variables. We use
the designations $^A$ and $^B$ for the longer and shorter periods,
respectively.
 
Fig.~\ref{f7} presents the light curves of the double-mode variables
phased with the periods of each mode after prewhitening with  the other
mode.
\section{Mode Identification}
Amplitudes generally yield no definitive clues for the identification of modes,
except that large amplitudes are more likely to occur in radial pulsations. 
Our identification of pulsation modes relies on the period ratios and
on the distribution of stars in the period-luminosity (P--L) plot.
\subsection{Amplitudes}
We observe amplitudes ranging from 0.016~mag. to 0.899~mag.  The
amplitude of V25 ($A_V$=0.899 mag) is one of the largest known among
all SX~Phe type variables. It is not likely that such an amplitude
arises in non-radial oscillations.  For most of the double-mode
variables the amplitude of the longer period oscillations is larger
than that for the shorter one.  An exception is V38 which has a larger
amplitude for the shorter period.  For this reason it is very likely
that in the double mode stars the oscillations with larger periods and
amplitudes are radial (\cite{Gilliland1998}).

In Fig.~\ref{f12} we present a color-amplitude relation for the stars
in our sample. Note that the larger amplitudes are exhibited by stars
close to the center of the instability strip.  The amplitudes of the
double mode stars tend to be smaller than the amplitudes of the single
mode stars, but a few single mode stars display very small amplitudes
as well.  Both effects, if real, are consistent with theoretical
expectations. However, the large scatter in Fig.~\ref{f12} makes any
detailed discussion of amplitude effects premature.
\subsection{Period Ratios}
The periods of the SX~Phe variables in M~55 span the range  0.033 to
0.136 days. We use $P_A$ for the longer periods and $P_B$ for the
shorter periods of the double mode variables.  Fig.~\ref{f8} presents
$P_B/P_A$ plotted against $P_A$ for the double-mode 
variables.  The $P_B/P_A$ ratio does not depend on the period of the
pulsations.  The weighted mean of $P_B/P_A$ for V31, V32, V33, V34,
V37, V38, and V42 is 0.975~$\pm$0.005. The period ratios of V35, V36,
V39 and V40 exhibit a larger scatter lying in the range $0.92 - 0.96$.
Since there are no radial modes spaced so closely in frequency, at
least one of the modes in our double-mode SX~Phe variables is
non-radial in origin. However we are unable to say with
assurance which of the two modes is radial, if any, using only period
information.
\subsection{Our Rosetta Stone: V41}
V41 is an exceptional case in that its  period ratio is extreme
compared to other double mode SX~Phe stars in M~55 (Fig.~\ref{f8}).
This period ratio helps us to identify its pulsation modes with some
confidence. The observed value of $0.807 \pm 0.009$ is close to the
first and second~overtone ratio for SX~Phe stars (0.801, see
\cite{Petersen1998} for a discussion).  For this reason we identify $P_A$
and $P_B$ with the first and second radial overtones, respectively.  In
Fig.~\ref{f9} we plot the period-luminosity relation for the principal
periods of all of the  stars in our sample. Except for V41 all of the
secondary periods of the double-mode stars lay close to their primary
periods and are not plotted to avoid confusion.  For V41 the secondary
period $P_B$ lies off of the general P--L relation, toward lower
periods. It is
consistent with our identification of $P_B$ with the second overtone.
This is true for all slopes of P--L relations discussed in
the literature, ranging from -3.3 to -3.7
(\cite{McNamara1995}, \cite{McNamara1997}). 
{\bf However, we caution that these results are extremely sensitive against
selection of the observational data. The latter paper claims 5-fold decrease
of scatter of $M_V$ without explainable improvement in the quality of the
observations.}
\section{The First Overtone P--L Relation}
In Fig. \ref{f9} the stars V20, V35, V36, V38 and V41 are marked with
filled symbols. These stars form a distinct branch away from the main
group of SX~Phe stars, shifted towards lower periods. Following our
identification of V41 as a first overtone pulsator we extend this
identification onto the whole group. 

Previous investigations have not revealed such a clear separation of the radial
modes of SX Phe stars in globular clusters.
These investigations have had to rely on small samples from different
clusters, and so relative distance errors and spatially variable
reddening both introduce significant scatter in the period-luminosity
diagram (\cite{McNamara1995}, \cite{Kaluzny1993}).

The dotted line in Fig.~\ref{f9} represents a  linear least squares
fit to the first overtone P--L relation:
\begin{eqnarray}
  <V> = -3.1 & log P_1 + & 12.32, \label{e.4} \\
      \pm0.6 &           & \pm0.03 \nonumber
\end{eqnarray}
with a standard residual of the fit of 0.05 mag.
\section{The Fundamental Mode P--L Relation for SX~Phe Stars}
\subsection{Slope}
We classify all remaining stars in Fig.~\ref{f9} (plotted with open
symbols) as SX~Phe stars pulsating in the fundamental mode.  The
continuous line in Fig.~\ref{f9} represents a linear least squares fit to
this P--L relation:
\begin{eqnarray}
<V> = -2.88 & log P_0 & + 13.09 \label{e.5} \\
    \pm0.17 &  ~      & \pm0.03 \nonumber
\end{eqnarray}
with a standard residual of the fit of 0.12 mag.

Our fundamental mode P--L relation is less steep than the overtone
relation.
{\bf However the relatively large error of the slope derived for
the first overtone P--L relation, does not reject the hypothesis of
equal slopes. This is} in agreement with the discussion by \cite{Nemec1994}.
This P--L relation for the fundamental mode stars exhibits a fair amount
of scatter.  The cause of this might be  misidentification among close
radial and non-radial modes. The average period ratio of $0.97$ in
bimodal stars  is consistent with a scatter of 0.03 in $\log P$ due to
mode misidentification.  In addition, some scatter is to be expected
from the finite width of the instability strip.

\cite{McNamara1995} derived a P--L relation with a slope $a=-3.3$ from a
compilation of cluster SX~Phe stars. This compilation relies on a smaller
and less
homogeneous data set than that presented here, and hence a realistic
estimate of the error of this latter value is expected to be large
compared to our error of $0.17$. Thus the \cite{McNamara1995} value for the
slope is marginally consistent with our value.   A comparison of these
results indicates the  degree of the external errors involved in P--L
relations for SX~Phe stars.

Our P--L slope of $a=-2.9$ is inconsistent with the
value $a=-3.7$ obtained by \cite{Petersen1998} from the parallaxes of
$\delta$~Scuti stars observed by HIPPARCOS. This is not surprising given
the observed scatter in the P--L relation for the HIPPARCOS stars. 
In addition, these calibrations do not take into account the fact that
SX~Phe itself is the star in the sample  with the shortest period 
and the lowest  metallicity at [Fe/H] = --1.37 (\cite{Hintz1998}).
The other 5 $\delta$~Scuti stars from the HIPPARCOS sample have high
metallicities ([Fe/H]$\simeq$0.0). \cite{Nemec1994} demonstrated
that SX Phe stars follow a period-luminosity-metallicity relation
with a coefficient of 0.32 for the [Fe/H] term, and so the slope of the
P--L relation from the HIPPARCOS stars will be  over-estimated.
{\bf Our P--L slope is also inconsistent with the value of --3.7 obtained
by \cite{McNamara1997} for the highly inhomogeneous sample 26~HADS, for which
P--L dependence was found indirectly, via many intermediate steps.}
\subsection{Zero point}
On the other hand the HIPPARCOS parallax of SX~Phe is
crucial for a determination of the zero point of the P--L relation for our
M~55 stars. The metallicity of SX~Phe is similar to 
 M~55 ([Fe/H]$=-1.54$, \cite{Rutledge1997}).
The parallax of SX~Phe ($\pi=12.91$~miliarcsec)
is well determined with a relative error $\sigma_\pi/\pi=0.06$.
The absolute magnitude of the SX~Phe, derived using the HIPPARCOS
parallax, is $M_V=2.88 \pm 0.13$~mag (\cite{Petersen1998}).
\cite{Oudmajier1998} determined that when the relative error of the
parallax is smaller than about 0.15,  the Lutz-Kelker correction
(\cite{Lutz1973}) accurately describes the probable shift in the
derived absolute magnitude.  In the case of SX~Phe, the Lutz-Kelker
correction is equal to $-0.02$~mag, so the corrected absolute
magnitude is $2.86$~mag.  This value, when combined with Eq.~(\ref{e.5})
for the fundamental mode period of SX~Phe of
$P_0=0.0550$~days (\cite{Petersen1998}), yields our final P--L
relation:
\begin{eqnarray}
M_V = -2.88 & log P_0 & - 0.77   \label{e.8} \\
    \pm0.17 &  ~      & \pm0.25 \nonumber
\end{eqnarray}
Using our calibration we determine the apparent distance modulus to
M~55 to be $(m-M)_V=13.86\pm0.25$ mag.  This result is consistent with
the apparent distance to M~55 determined from the analysis of RR~Lyrae
stars in M~55 by \cite{Olech1999}.
\section{The Period--Color Relation}
$\delta$ Scuti and SX~Phe stars close to the red border of the
instability strip have periods significantly longer than the periods of
stars at the  center of the strip (\cite{Pamiatnykh2000}).
The period-color ($log P / (V-I)$) dependence for $\delta$~Scuti stars
from the Galactic Bulge was described  by {\cite{McNamara2000}.
A linear least squares fit to our data, presented in Fig.~\ref{f11} 
yields the following relation:
\begin{eqnarray}
  <B-V> = 0.15 & log P & + 0.543 \label{e.6} \\
       \pm0.05 &       & \pm0.008 \nonumber
\end{eqnarray}
The standard residuals from the fit amount to $0.05$~mag.
Note that the {\bf star} towards
lower right in Fig.~\ref{f9} {\bf is V21, which is found} at the extreme
red border of the instability strip {\bf (Fig.~\ref{f4}). }
Any attempt to account for this by including a color term in the P--L
relation failed to improve the quality of the fit. 
\section{Conclusions}
SX~Phe type variables seem to be good distance indicators. Although
their luminosities are too low for investigations in distant galaxies, 
they are bright enough to be observed in nearby galaxies.
{\bf The largest number of SX~Phe variables in one globular cluster
was found in $\omega$~Cen (\cite{Kaluzny1996}, \cite{Kaluzny1997a}),
but due to its varying
metallicity this cluster is not suitable for distance calibration.}
In M~55 we discovered 
the richest population of SX~Phe among the rest of globular clusters. 
{\bf M~55 is thought to be chemically homogeneous (\cite{Richter1999}).}
This enabled a separation of the fundamental and first overtone stars and
an estimate of the errors caused by
misidentification of nearby radial non-radial frequencies. In this way
we obtained a reliable slope of the P--L relation for the
fundamental mode stars. Combined with the HIPPARCOS parallax for SX~Phe
itself, we obtain an improved P--L relation (Eq.~\ref{e.8}).
Despite being based on just one star, our zero point should be reliable
as HIPPARCOS parallax of SX~Phe has an error of $6$~percent and
metallicities of SX~Phe and of M~55 are as close as $-1.37$ and $-1.54$.
Using our revised P--L relation for SX~Phe stars we measure the apparent
distance to M~55 to be $(m-M)_V=13.86\pm0.25$~mag. 
\begin{acknowledgements}
We would like to thank Alosha Pamiatnykh and Wojciech Dziembowski for their
enlightening comments.
JK and  WK were supported by the KBN grant 2P03D.003.17.
WP was supported by the KBN grant 2P03D.010.15.
JK was supported also by NSF grant AST 9819787 to B. Paczy{\'n}ski.
IBT and WK were supported by NSF grant AST-9819786.
ASC was supported by the KBN grant No. 2P03D 018 18.
\end{acknowledgements}
\clearpage
%
%
\begin{table}
\caption{List of single-mode SX~Phe variables in the field of M~55 \label{t1} }
\begin{tabular}{lccccccc}
\hline
star & R.A.(J2000.0) & Dec.(J2000) & P & $<V>$ & $<B>$ & $<$$B$$>-<$$V$$>$ & $A_V$ \\
    & hh:mm:sec      & deg:':"	& [days] & & & & \\
\hline 
V16 & 19:40:09.20 & -30:56:42.04 & 0.0534204(8) & 16.94 & 17.32 & 0.38 & 0.016 \\
V17 & 19:40:11.33 & -30:59:25.06 & 0.0412615(3) & 17.18 & 17.43 & 0.25 & 0.049 \\
V18 & 19:40:06.87 & -30:56:32.12 & 0.0465555(4) & 16.98 & 17.32 & 0.34 & 0.029 \\
V19 & 19:39:57.67 & -30:57:01.31 & 0.0382367(2) & 17.27 & 17.64 & 0.37 & 0.033 \\
V20 & 19:39:54.95 & -30:58:21.25 & 0.0332118(2) & 17.04 & 17.34 & 0.30 & 0.102 \\
V21 & 19:39:58.27 & -30:59:06.05 & 0.1355924(2) & 15.76 & 16.19 & 0.43 & 0.036 \\
V22 & 19:40:07.80 & -31:00:12.60 & 0.0456394(1) & 16.81 & 17.17 & 0.36 & 0.337 \\
V23 & 19:39:51.82 & -30:55:52.83 & 0.0413989(3) & 17.22 & 17.58 & 0.36 & 0.050 \\
V24 & 19:39:45.49 & -30:56:02.68 & 0.0418206(5) & 17.06 & 17.40 & 0.34 & 0.026 \\
V25 & 19:39:51.55 & -30:56:21.27 & 0.0985309(1) & 15.88 & 16.23 & 0.35 & 0.899 \\
V26 & 19:39:47.06 & -30:57:33.98 & 0.0820096(2) & 16.11 & 16.51 & 0.40 & 0.173 \\
V27 & 19:39:54.05 & -30:58:07.46 & 0.0410265(5) & 17.09 & 17.45 & 0.36 & 0.029 \\
V28* & 19:40:15.04 & -31:05:15.03 & 0.0537630(6) & 20.61 & 20.92 & 0.31 & 0.260 \\
V29* & 19:39:42.58 & -30:55:58.34 & 0.0343115(2) & 20.71 & 20.92 & 0.21 & 0.295 \\
V30* & 19:39:41.02 & -30:50:25.23 & 0.0563464(5) & 20.35 & 20.68 & 0.33 & 0.258 \\
\hline
\end{tabular}
* Probable member of the Sagittarius dwarf galaxy.
\end{table}
%
%
%
\begin{table}
\caption{Fourier parameters of single-mode SX~Phe variables in M~55.
See Eqs.~(\ref{e.2}) and (\ref{e.3}) for the definition. \label{t2} }
\begin{tabular}{cccccc}
\hline
star   &  $A_1$  & $R_{21}$ & ${\phi}_{21}$ & $R_{31}$ & ${\phi}_{31}$ \\
\hline 
V16 & 0.0079 {\footnotesize ${\pm}$0.0005} & - & - & - & - \\
V17 & 0.0246 {\footnotesize ${\pm}$0.0009} & - & - & - & - \\
V18 & 0.0144 {\footnotesize ${\pm}$0.0006} & - & - & - & - \\
V19 & 0.0166 {\footnotesize ${\pm}$0.0006} & - & - & - & - \\
V20 & 0.0500 {\footnotesize ${\pm}$0.0016} & 0.126 {\footnotesize ${\pm}$0.033} & 2.33 {\footnotesize ${\pm}$0.27} & - & - \\
V21 & 0.0178 {\footnotesize ${\pm}$0.0004} & 0.125 {\footnotesize ${\pm}$0.021} & 2.55 {\footnotesize ${\pm}$0.17} & - & - \\
V22 & 0.1609 {\footnotesize ${\pm}$0.0023} & 0.250 {\footnotesize ${\pm}$0.015} & 2.06 {\footnotesize ${\pm}$0.06} & - & - \\
V23 & 0.0252 {\footnotesize ${\pm}$0.0008} & - & - & - & - \\
V24 & 0.0130 {\footnotesize ${\pm}$0.0007} & - & - & - & - \\
V25 & 0.3454 {\footnotesize ${\pm}$0.0008} & 0.480 {\footnotesize ${\pm}$0.003} & 2.19 {\footnotesize ${\pm}$0.01} & 0.240 {\footnotesize ${\pm}$0.003} & 4.50 {\footnotesize ${\pm}$0.01} \\
V26 & 0.0839 {\footnotesize ${\pm}$0.0006} & 0.221 {\footnotesize ${\pm}$0.007} & 2.16 {\footnotesize ${\pm}$0.03} & 0.034 {\footnotesize ${\pm}$0.007} & 4.66 {\footnotesize ${\pm}$0.21} \\
V27 & 0.0143 {\footnotesize ${\pm}$0.0009} & - & - & - & - \\
V28 & 0.1276 {\footnotesize ${\pm}$0.0047} & 0.190 {\footnotesize ${\pm}$0.038} & 2.01 {\footnotesize ${\pm}$0.21} & - & - \\
V29 & 0.1477 {\footnotesize ${\pm}$0.0063} & - & - & - & -\\
V30 & 0.1191 {\footnotesize ${\pm}$0.0042} & 0.321 {\footnotesize ${\pm}$0.036} & 2.13 {\footnotesize ${\pm}$0.13} & 0.132 {\footnotesize ${\pm}$0.034} & 4.59 {\footnotesize ${\pm}$0.28} \\
\hline
\end{tabular}
\end{table}
\clearpage
%
%
\begin{table}
\caption{List of double-mode SX~Phe variables in M~55\label{t3} }
\begin{tabular}{lcccccccc}
\hline
star & R.A.(J2000.0) & Dec.(J2000.) & $P_A$ & $P_B$ & $<V>$ & $<B>$ & $<$$B$$>-<$$V$$>$ & $A_V$($P_A$) \\
    & hh:mm:sec      & deg:':"	& [days] & [days] & & & & \\
\hline
V31 & 19:40:00.99 & -30:57:56.53 & 0.0388471(2) & 0.0382042(5) & 17.23 & 17.60 & 0.37 & 0.041 \\
V32 & 19:39:58.14 & -30:58:32.66 & 0.0414874(2) & 0.0405483(5) & 16.92 & 17.28 & 0.36 & 0.097 \\
V33 & 19:39:54.56 & -30:59:57.88 & 0.0593067(3) & 0.0573473(4) & 16.40 & 16.74 & 0.34 & 0.054 \\
V34 & 19:40:01.02 & -31:00:37.93 & 0.0370203(3) & 0.0360939(6) & 17.23 & 17.54 & 0.31 & 0.029 \\
V35 & 19:39:50.37 & -30:55:12.41 & 0.0486858(2) & 0.0452695(7) & 16.57 & 16.91 & 0.34 & 0.070 \\
V36 & 19:39:48.56 & -30:56:45.04 & 0.0393958(2) & 0.0373768(4) & 16.74 & 17.05 & 0.31 & 0.067 \\
V37 & 19:39:49.87 & -30:57:42.53 & 0.0437976(3) & 0.0428239(7) & 16.96 & 17.26 & 0.30 & 0.051 \\
V38 & 19:39:58.86 & -30:58:14.78 & 0.0391971(5) & 0.0381747(3) & 16.69 & 17.04 & 0.35 & 0.044* \\
V39 & 19:40:11.99 & -31:02:04.48 & 0.0358151(3) & 0.0341535(5) & 17.21 & 17.52 & 0.31 & 0.034 \\
V40 & 19:40:01.90 & -30:55:38.20 & 0.0369762(4) & 0.0346515(7) & 17.20 & 17.56 & 0.35 & 0.028 \\
V41 & 19:40:02.95 & -30:58:28.34 & 0.0451669(2) & 0.0364552(3) & 16.53 & 16.83 & 0.30 & 0.106 \\
V42 & 19:39:58.61 & -30:57:23.93 & 0.0366655(3) & 0.0356382(5) & 17.16 & 17.52 & 0.36 & 0.053 \\
\hline
\end{tabular}
* Amplitude of $P_B$ pulsations (see text)
\end{table}
%
%
\begin{table}
\caption{Fourier parameters of double-mode SX~Phe variables in M~55.
See Eqs.~(\ref{e.2}) and (\ref{e.3}) for the definition. \label{t4} }
\begin{tabular}{ccccc}
\hline
star   & $A_{1}^{A}$ & $R_{21}^{A}$ & ${\phi}_{21}^{A}$ & $A_{1}^{B}$\\
\hline
V31 & 0.0204 {\footnotesize ${\pm}$0.0006} & 0.074 {\footnotesize ${\pm}$0.029} & 1.19 {\footnotesize ${\pm}$0.40} & 0.0093 {\footnotesize ${\pm}$0.0006} \\
V32 & 0.0478 {\footnotesize ${\pm}$0.0011} & 0.143 {\footnotesize ${\pm}$0.023} & 2.20 {\footnotesize ${\pm}$0.17} & 0.0194 {\footnotesize ${\pm}$0.0011} \\
V33 & 0.0271 {\footnotesize ${\pm}$0.0005} & -           & -         & 0.0181 {\footnotesize ${\pm}$0.0005} \\
V34 & 0.0144 {\footnotesize ${\pm}$0.0006} & -           & -         & 0.0060 {\footnotesize ${\pm}$0.0006} \\
V35 & 0.0349 {\footnotesize ${\pm}$0.0007} & -           & -         & 0.0099 {\footnotesize ${\pm}$0.0008} \\
V36 & 0.0333 {\footnotesize ${\pm}$0.0007} & -           & -         & 0.0119 {\footnotesize ${\pm}$0.0007} \\
V37 & 0.0254 {\footnotesize ${\pm}$0.0009} & -           & -         & 0.0106 {\footnotesize ${\pm}$0.0009} \\
V38 & 0.0122 {\footnotesize ${\pm}$0.0008} & -           & -         & 0.0218 {\footnotesize ${\pm}$0.0008} \\
V39 & 0.0170 {\footnotesize ${\pm}$0.0006} & -           & -         & 0.0080 {\footnotesize ${\pm}$0.0006} \\
V40 & 0.0141 {\footnotesize ${\pm}$0.0006} & -           & -         & 0.0068 {\footnotesize ${\pm}$0.0006} \\
V41 & 0.0532 {\footnotesize ${\pm}$0.0008} & -           & -         & 0.0243 {\footnotesize ${\pm}$0.0008} \\
V42 & 0.0263 {\footnotesize ${\pm}$0.0010} & -           & -         & 0.0171 {\footnotesize ${\pm}$0.0010} \\
\hline
\end{tabular}
\end{table}
%
%
\clearpage
\begin{figure*}
\resizebox{\hsize}{!}{\includegraphics{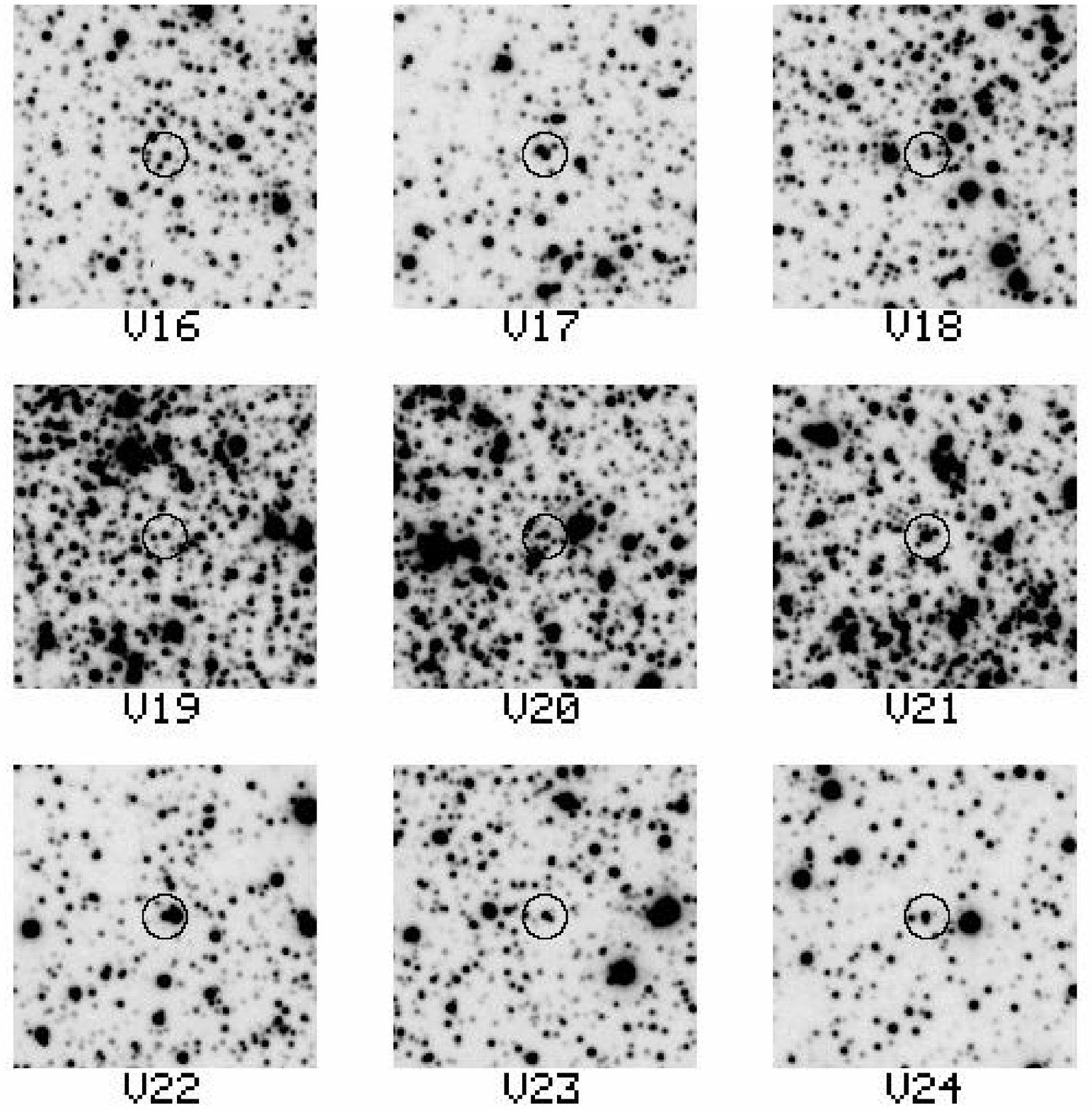}}
\caption{Finding charts for SX~Phe variables in M~55. Part I - variables
V16 - V24. Each chart is 1~arcmin on a side, with north to the top and east to
the left.}
\label{f1}
\end{figure*}
%
%
\clearpage
\begin{figure*}
\resizebox{\hsize}{!}{\includegraphics{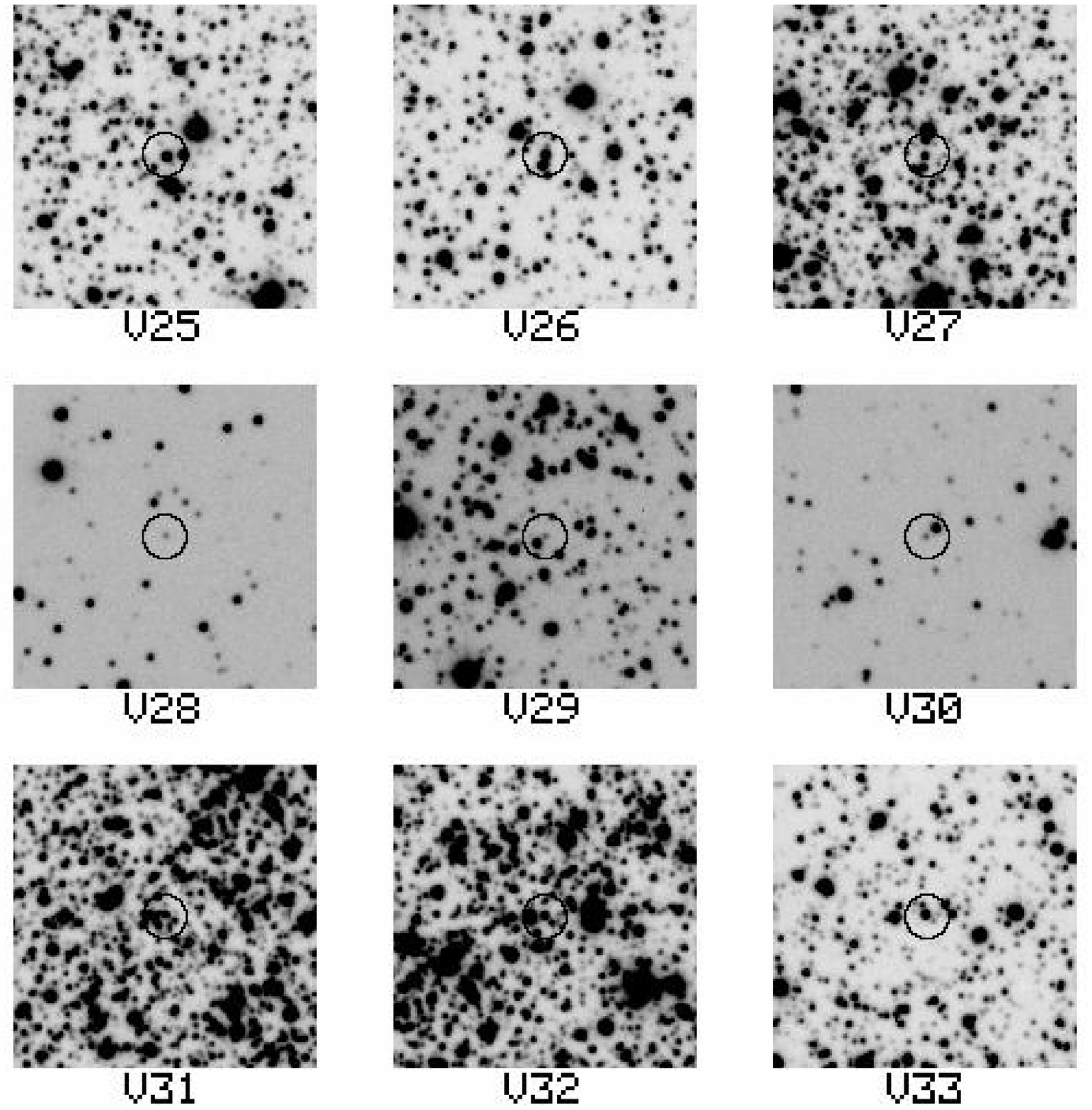}}
\caption{Finding charts for SX~Phe variables in M~55. Part II - variables
V25 - V33. Each chart is 1~arcmin on a side, with north to the top and east to
the left.
\label{f2}}
\end{figure*}
%
%
\clearpage
\begin{figure*}
\resizebox{\hsize}{!}{\includegraphics{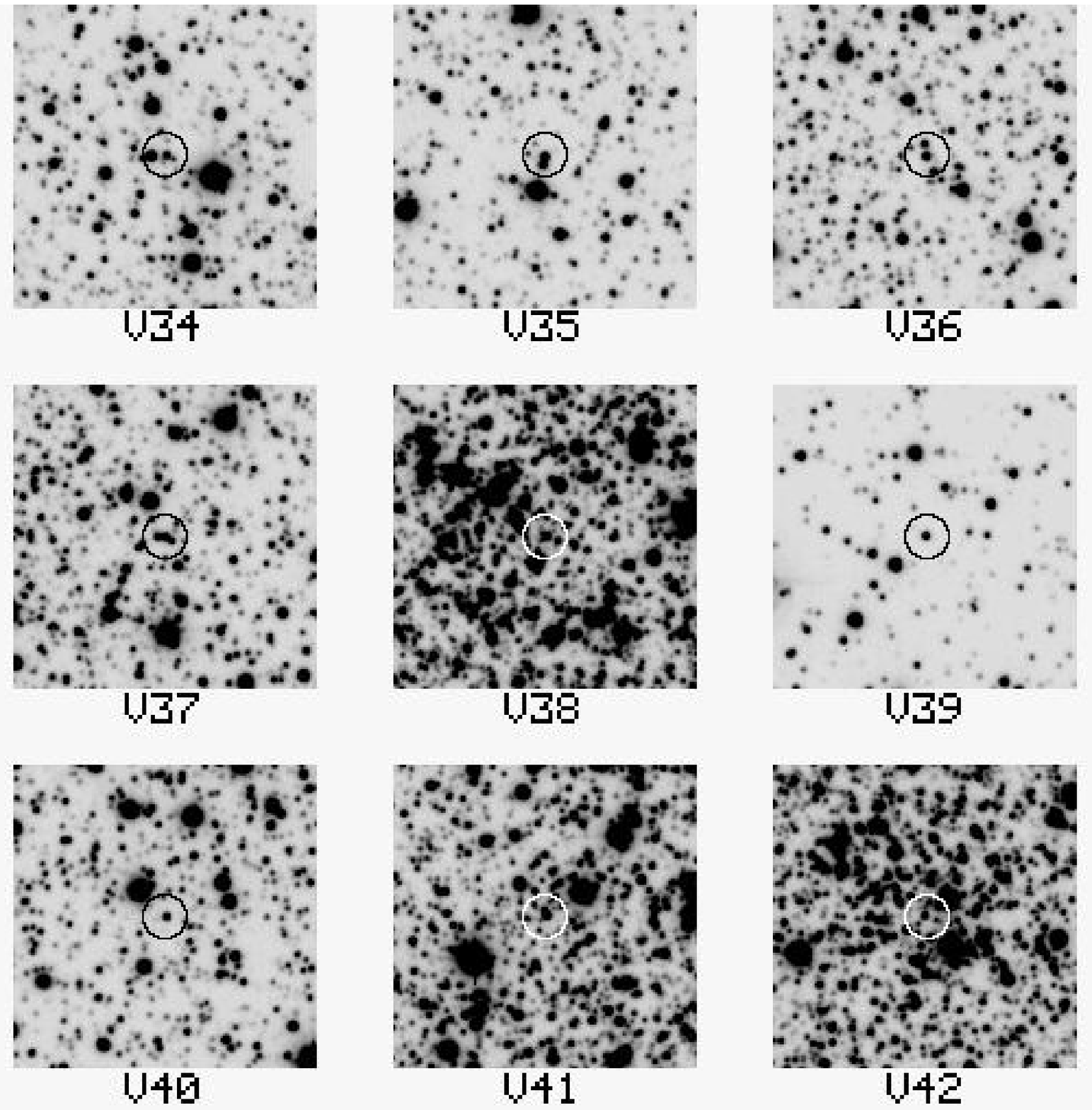}}
\caption{Finding charts for SX~Phe variables in M~55. Part III - variables
V34 - V42. Each chart is 1~arcmin on a side, with north to the top and east to
the left.
\label{f3}}
\end{figure*}
%
%
\clearpage
\begin{figure}
\resizebox{\hsize}{!}{\includegraphics{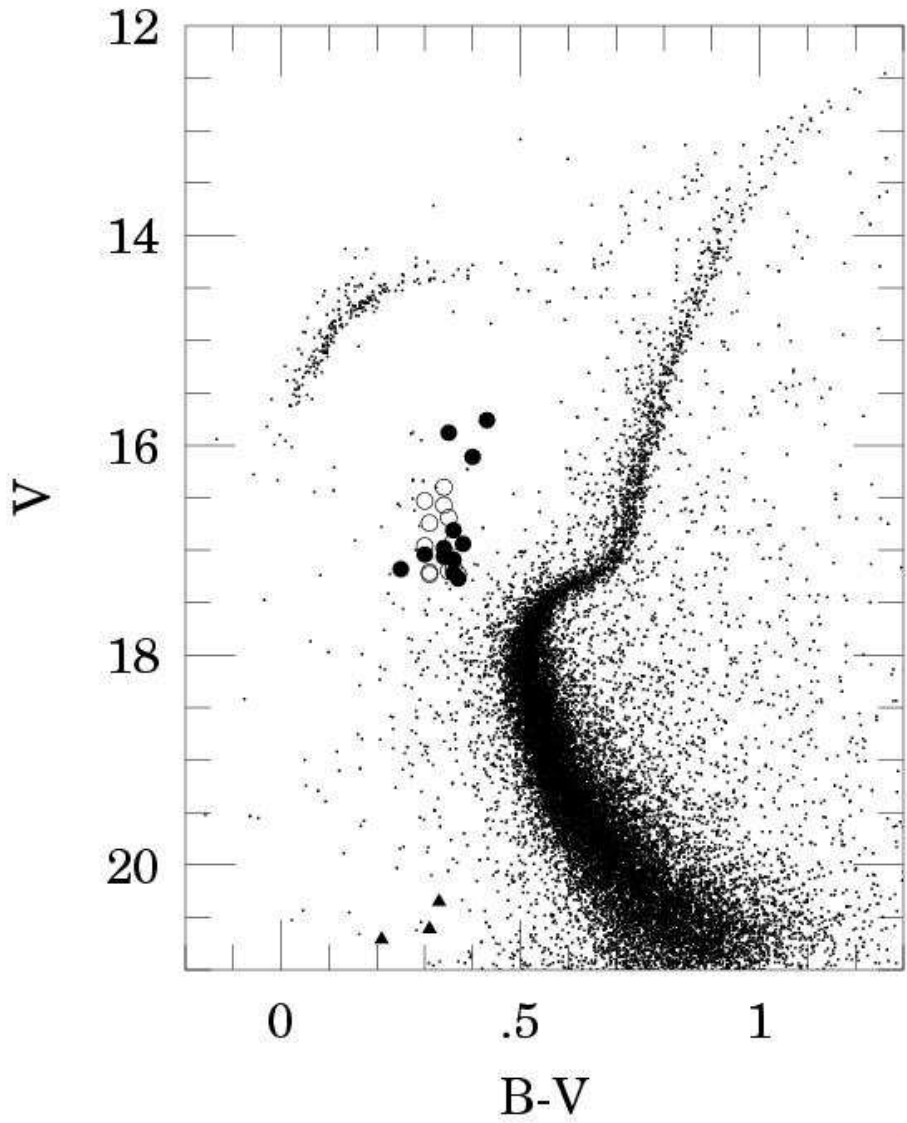}}
\caption{The color-magnitude diagram of the globular cluster M~55. Filled
circles denote single-mode SX~Phe variables, open circles denote
double-mode SX~Phe variables, filled triangles denote SX~Phe variables
from the Sagittarius dwarf galaxy.
\label{f4} }
\end{figure}
%
%
\clearpage
\begin{figure}
\resizebox{\hsize}{!}{\includegraphics{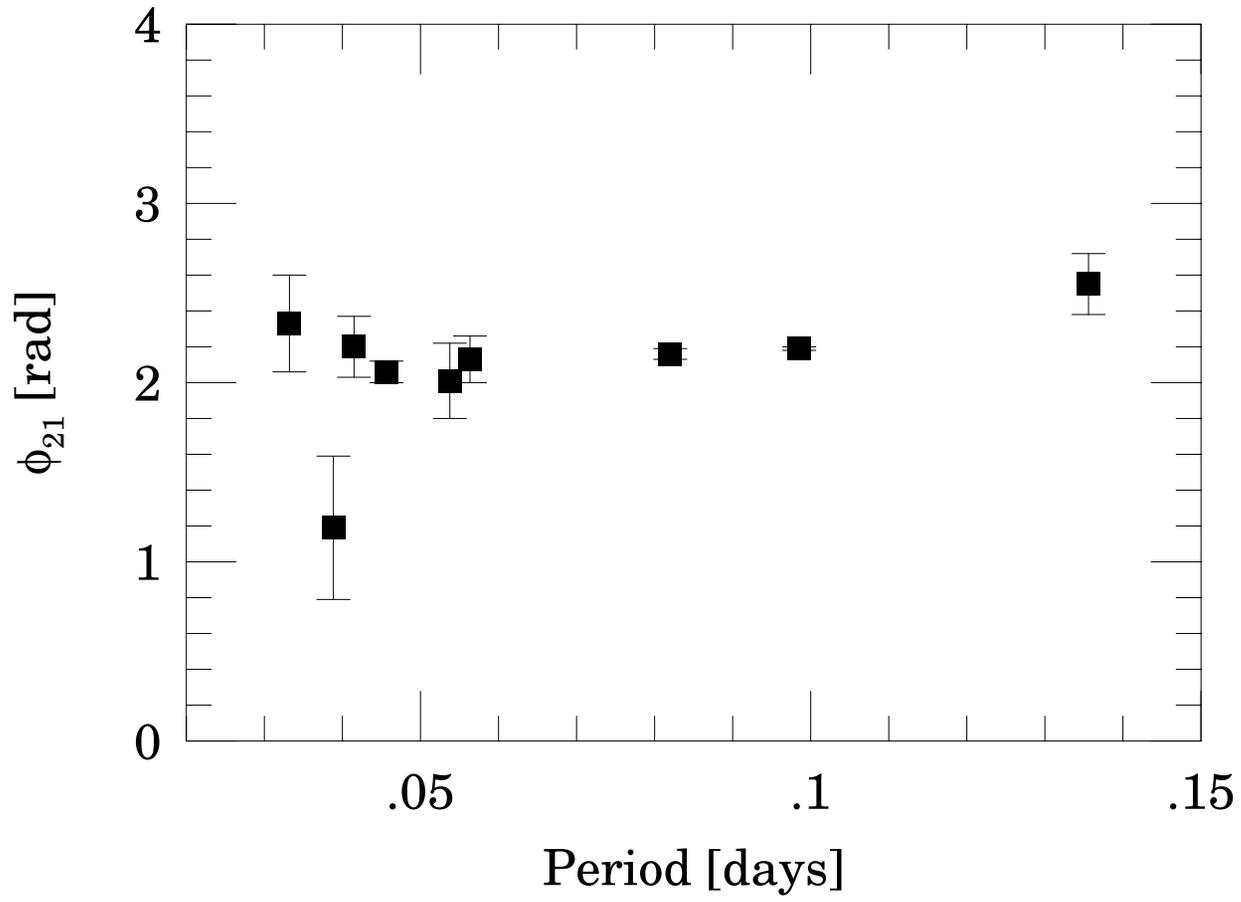}}
\caption{$\phi_{21}$-period relation for SX~Phe stars in M~55.
\label{f5} }
\end{figure}
%
%
\clearpage
\begin{figure*}
\resizebox{\hsize}{!}{\includegraphics{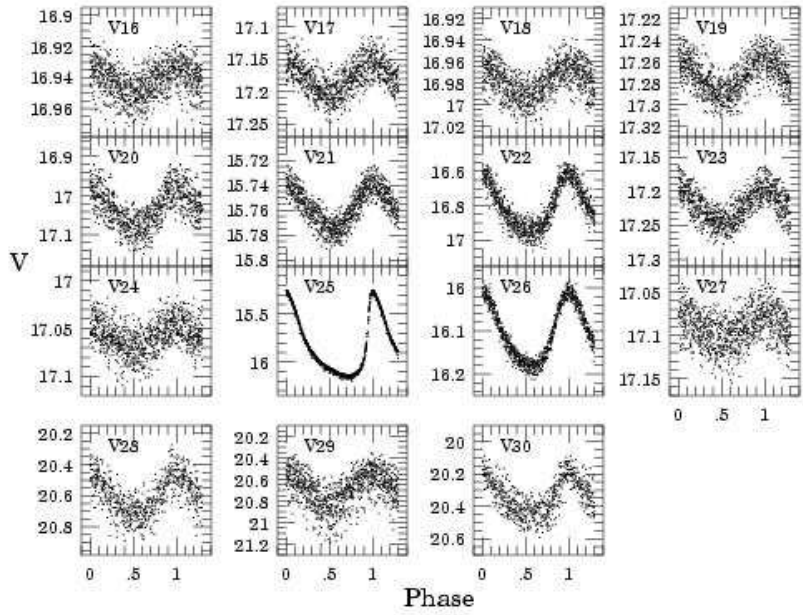}}
\caption{$V$-band light curves for single-mode SX~Phe variables observed in
the field of M~55
\label{f6}}
\end{figure*}
%
%
\clearpage
\begin{figure*}
\resizebox{\hsize}{!}{\includegraphics{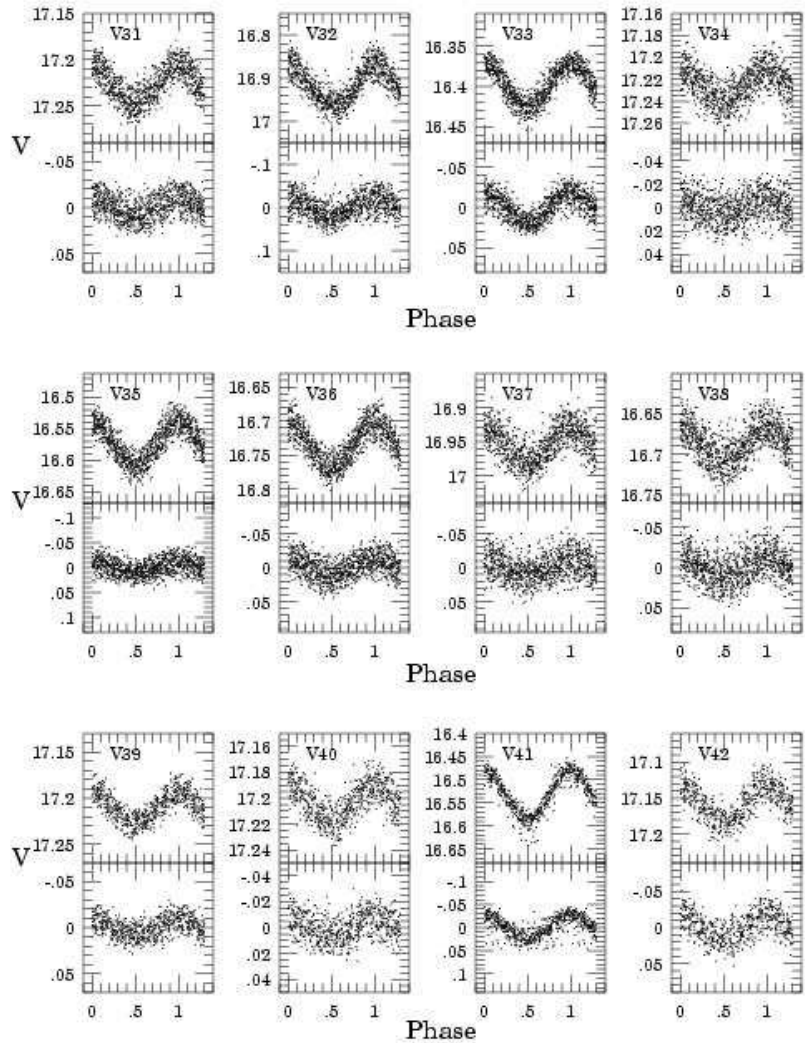}}
\caption{$V$-band light curves for double-mode SX~Phe variables observed in
the field of M~55. The upper panels show light curves phased with longer
period after subtraction of variability with the shorter period
approximated by the Fourier series.
The lower panels show light curves phased with shorter period after 
subtraction of variability with the longer period
approximated by the Fourier series.
\label{f7} }
\end{figure*}
%
%
\clearpage
\begin{figure}
\resizebox{\hsize}{!}{\includegraphics{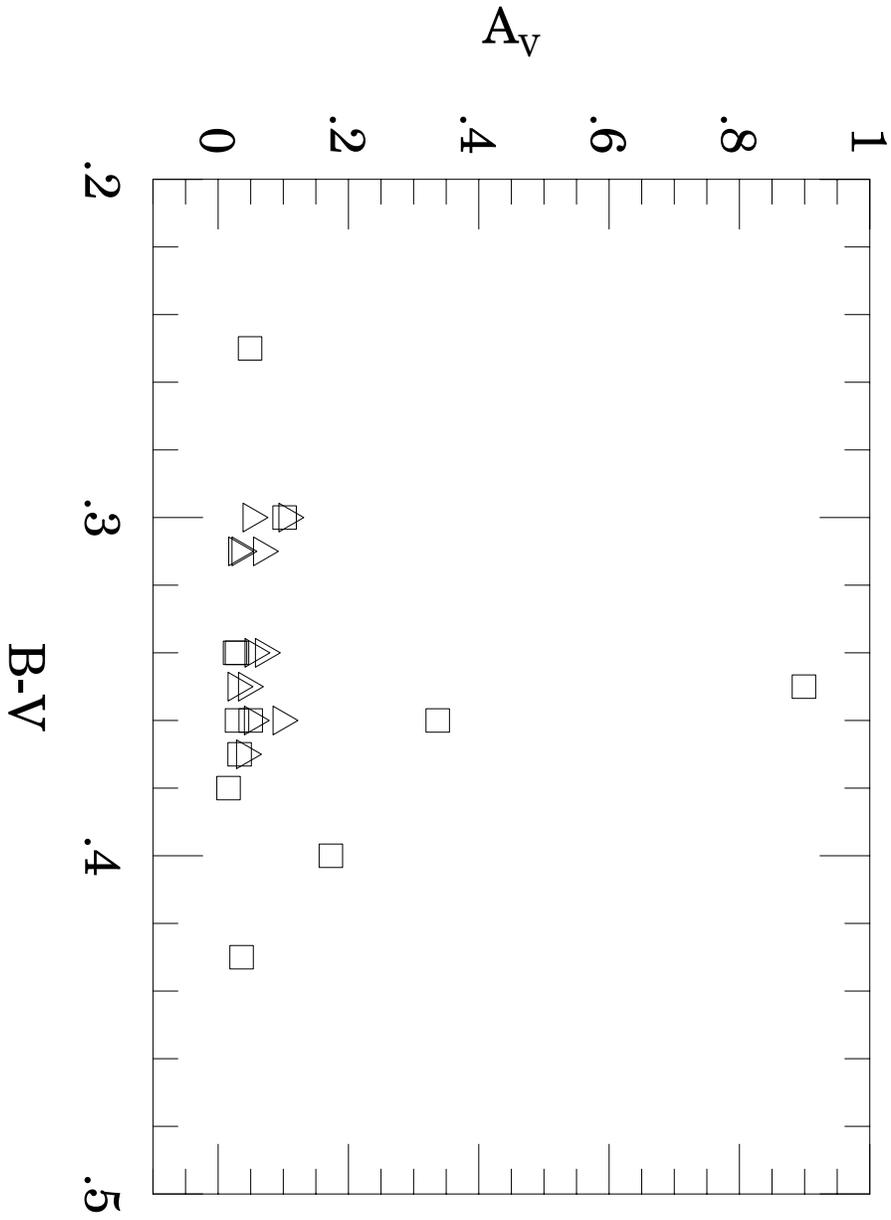}}
\caption{The color-amplitude relation for SX~Phe stars in M~55. Squares
denote single mode pulsators, triangles denote double mode pulsators.
\label{f12} }
\end{figure}
%
%
%
\clearpage
\begin{figure}
\resizebox{\hsize}{!}{\includegraphics{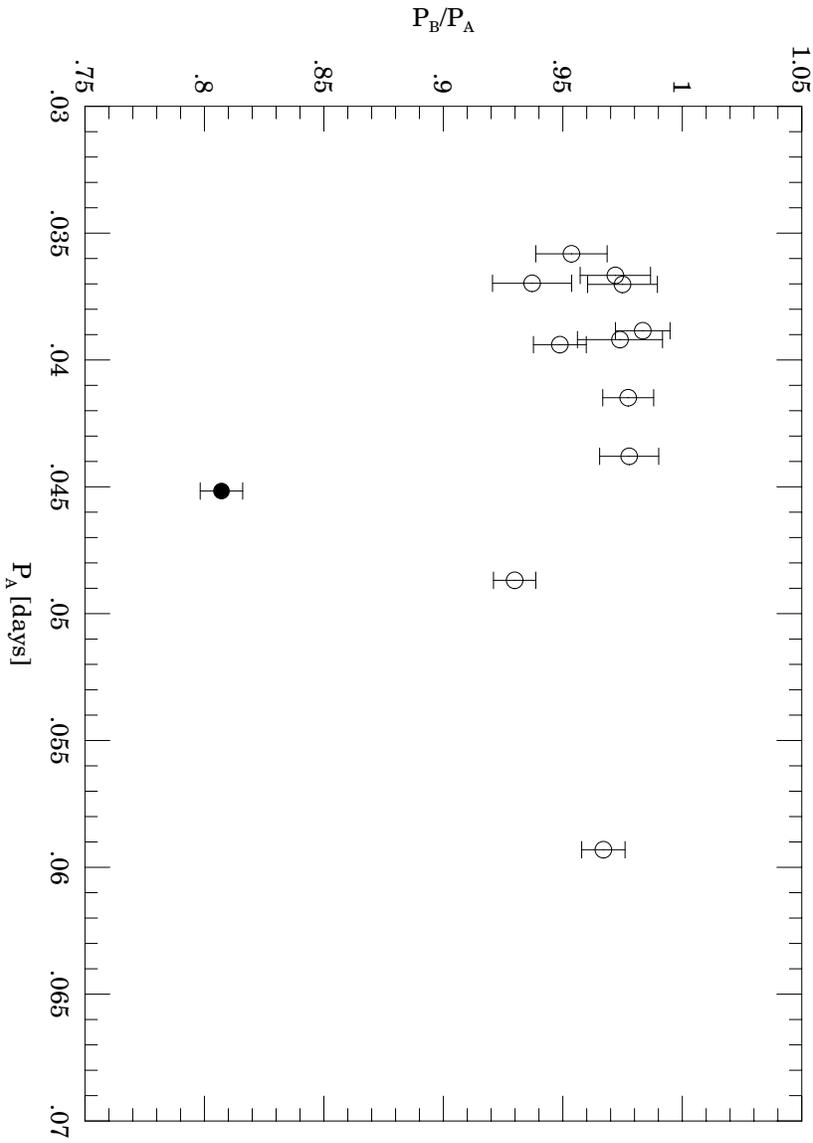}}
\caption{Period ratio ($P_B/P_A$) versus $P_A$ relation for
double mode SX~Phe variables in M~55.
The open symbols represent the stars with the period ratios
$0.9<P_B/P_A<1.0$ - at least one of the modes must be non-radial. 
The filled circle denotes V41 with the period ratio equal to 0.807 - 
characteristic for pulsations in the first and the second radial
overtones.
\label{f8} }
\end{figure}
%
%
\clearpage
\begin{figure*}
\resizebox{\hsize}{!}{\includegraphics{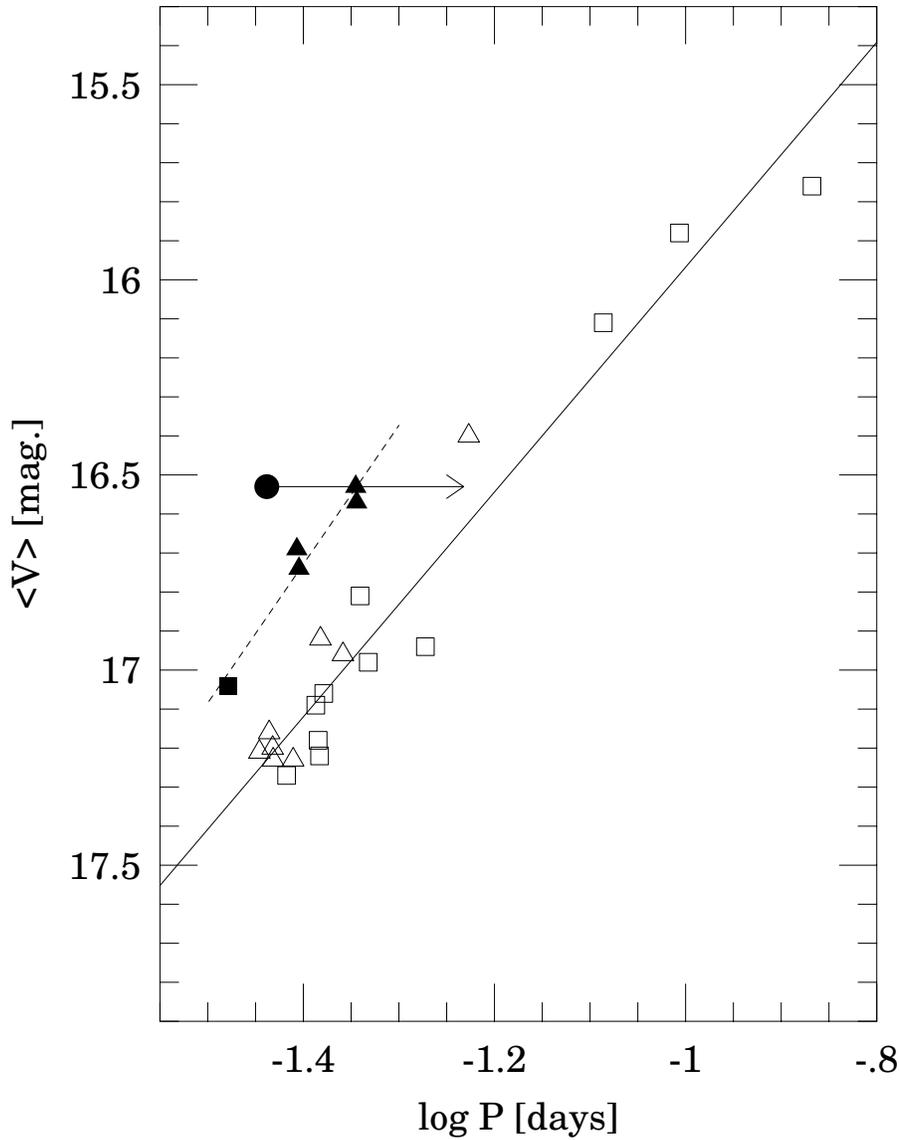}}
\caption{The period-luminosity dependence for SX~Phe stars in M~55.
The squares denote single-mode variables, the triangles denote points 
corresponding to longer periods of double-mode variables,
the circle denotes the point corresponding to the second overtone period
of V41. The solid line represents linear least squares fit
to the points marked with open symbols (fundamental mode pulsators).
The dashed line represents linear least squares fit to the points marked
with filled symbols (first overtone pulsators).
The horizontal line represents the period 
shifts of V41: between the first and the second overtones, and to the
fundamental mode.
\label{f9} }
\end{figure*}
%
%
%
\clearpage
\begin{figure}
\resizebox{\hsize}{!}{\includegraphics{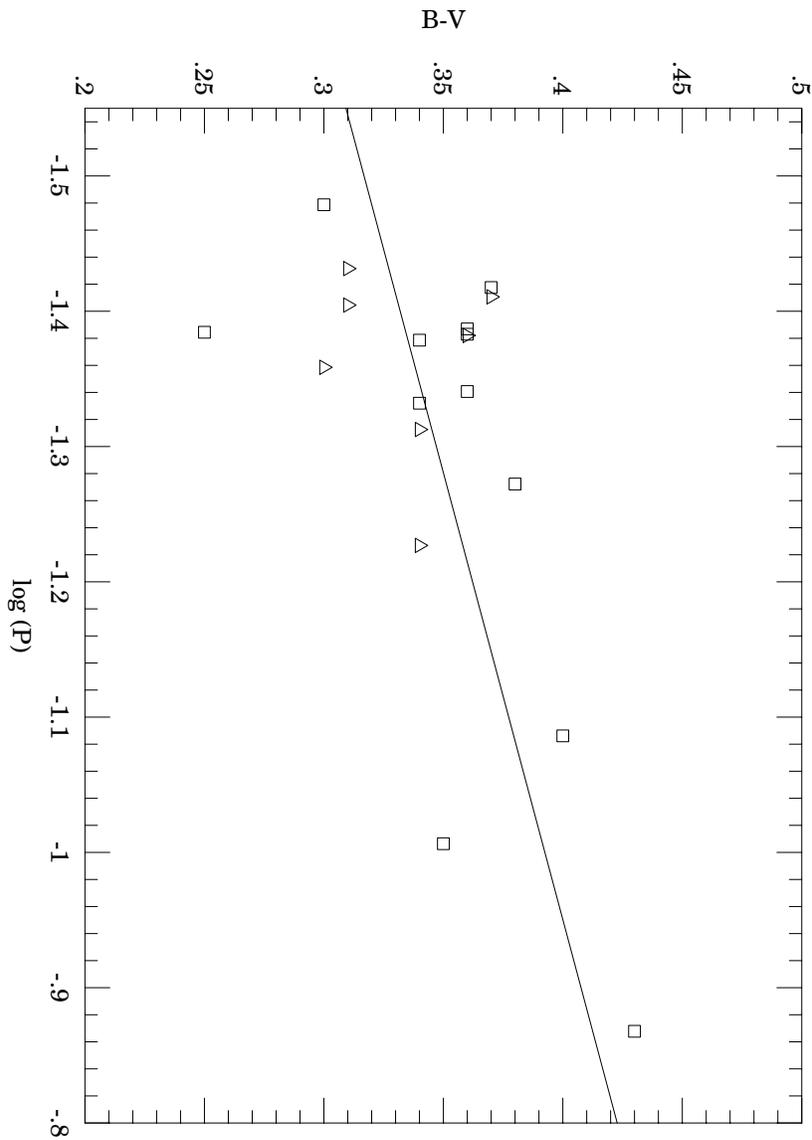}}
\caption{The period-color dependence for SX~Phe stars in M~55.
The squares denote single-mode variables, the triangles denote points 
corresponding to longer periods of double-mode variables.
The solid line represents linear least squares fit
drawn through the plotted points.
\label{f11} }
\end{figure}
\end{document}